\documentclass[twocolumn,superscriptaddress,preprintnumbers,floatfix]{revtex4}
\usepackage{graphicx}
\usepackage{dcolumn}

\usepackage[usenames]{color}

\graphicspath{{Fig/}}

\begin{document}
\sloppy

\title{Optical spectra obtained from amorphous films of rubrene: Evidence for
predominance of twisted isomer}

\author{M.~Kytka}

\affiliation{Institut f\"ur Angewandte Physik, Universit\"at
  T\"ubingen, Auf der Morgenstelle 10, 72076 T\"ubingen, Germany}

\affiliation{Department of Microelectronics, Slovak University of
  Technology and International Laser Center, Ilkovi\v cova 3, 812 19
  Bratislava, Slovakia}

\author{L.~Gisslen}

\affiliation{Walter Schottky Institut, Technische Universit\"at M\"unchen, Am
  Coulombwall 3, 85748 Garching, Germany}

\author{A.~Gerlach}

\affiliation{Institut f\"ur Angewandte Physik, Universit\"at
  T\"ubingen, Auf der Morgenstelle 10, 72076 T\"ubingen, Germany}

\author{U.~Heinemeyer}

\affiliation{Institut f\"ur Angewandte Physik, Universit\"at
  T\"ubingen, Auf der Morgenstelle 10, 72076 T\"ubingen, Germany}

\author{J.~Kov\'a\v c}

\affiliation{Department of Microelectronics, Slovak University of
  Technology and International Laser Center, Ilkovi\v cova 3, 812 19
  Bratislava, Slovakia}

\author{R.~Scholz}

\affiliation{Walter Schottky Institut, Technische Universit\"at M\"unchen, Am
  Coulombwall 3, 85748 Garching, Germany}

\author{F.~Schreiber}
\email{frank.schreiber@uni-tuebingen.de}

\affiliation{Institut f\"ur Angewandte Physik, Universit\"at
  T\"ubingen, Auf der Morgenstelle 10, 72076 T\"ubingen, Germany}

\date{\today}

%
%

\begin{abstract}
  In order to investigate the optical properties of rubrene we study
  the vibronic progression of the first absorption band (lowest
  $\pi\!\rightarrow\!\pi^*$ transition). We analyze the dielectric
  function $\varepsilon_2$ of rubrene in solution and thin films using
  the displaced harmonic oscillator model and derive all relevant
  parameters of the vibronic progression.
  The findings are supplemented by density functional calculations
  using B3LYP hybrid functionals. Our theoretical results for the
  molecule in two different conformations, i.e.\ with a twisted or
  planar tetracene backbone, are in very good agreement with the
  experimental data obtained for rubrene in solution and thin films.
  Moreover, a simulation based on the monomer spectrum and the
  calculated transition energies of the two conformations indicates
  that the thin film spectrum of rubrene is dominated by the twisted
  isomer.
\end{abstract}
\maketitle

\section{Introduction}

\begin{figure}[htbp]
  \centering 
  \includegraphics[width=0.7\columnwidth]{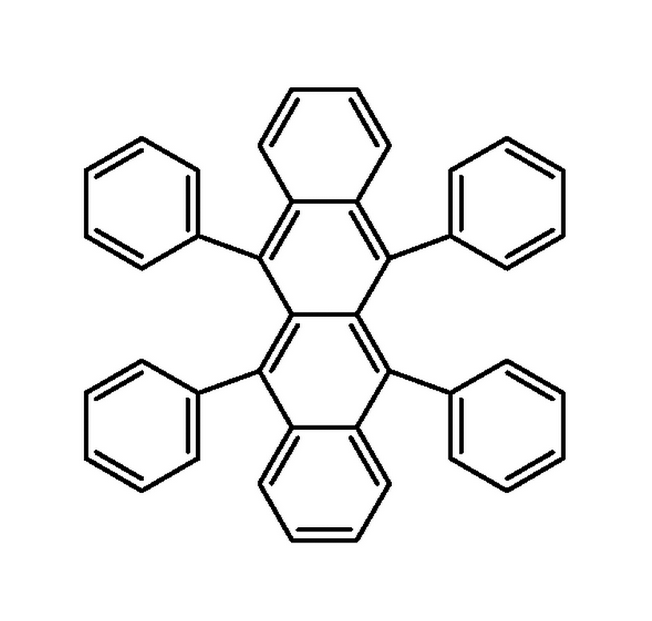}
  \caption{Structure of the rubrene molecule (C$_{42}$H$_{28}$) which
    consists of a tetracene backbone (TBB) and four attached phenyl
    rings. Due to the different orientation of the phenyl rings
    and TBB the $\pi$-conjugated systems of both parts are strongly
    decoupled.}
  \label{fig:structure}
\end{figure}

Organic semiconductors have become very promising and actively
investigated materials
\cite{Bruetting_bk,Witte_pssa08,xGerlach_pssa08}. Due to their specific
optical properties organic molecules are particularly interesting for
various device applications such as light emitting diodes, field
effect transistors and solar cells \cite{Park_apl07-1,
  Park_apl07-2,Hsu_apl07}.
Different studies have demonstrated that the optical properties of
molecular solids are strongly influenced by their
structure. Crystalline materials as for example pentacene
\cite{Hinderhofer_jcp07} and diindenoperylene \cite{Heinemeyer_prb08}
thin films exhibit complex and generally anisotropic absorption
spectra which are often influenced by characteristic charge transfer
excitations. Amorphous materials with extended $\pi$-conjugated
electron systems, as for example rubrene (Fig.~\ref{fig:structure}),
show optical spectra which are seemingly simpler to understand
\cite{Kaefer_pccp05,Kaefer_prl05,Kowarik_pccp06}.
However, when considered in more detail the growth of amorphous
materials may present its own complications. For example, studies of
rubrene thin films on silicon uncovered a surprising growth behavior:
Using X-ray reflectivity rubrene thin films are found to be relatively
smooth and amorphous over a large thickness range
\cite{Kowarik_pccp06}. Yet, an anomalous roughness evolution of the
films is observed, i.e.\ a decrease of the film roughness at the
beginning (below $\sim$20\,nm) and a subsequent increase of the
roughness at a later stage.  Recent near edge X-ray absorption fine
structure (NEXAFS) results indicate the existence of different rubrene
isomers on SiO$_2$ and Au(111) \cite{Kaefer_prl05}. At the early stage
of growth both species were identified, i.e.\ the rubrene with the
twisted and planar tetracene backbone, followed by the planar isomer
only.

In the present study we focus on the optical properties of rubrene
molecules in solution and amorphous thin films. In particular, we
investigate whether the different isomers can be identified in the
optical spectra.  For this purpose we calculate the optical response
of the different isomers of rubrene with time-dependent density
functional theory (TD-DFT). The deformation in the relaxed excited
geometry of each isomer found with TD-DFT is projected onto its
internal vibrations. This allows an assignment of the observed
sub-bands in the absorption spectra to an effective vibrational mode
which is an average over the various internal vibrations elongated in
the relaxed excited geometry. Furthermore, we use the calculations to
assess the contribution of the different rubrene isomers to the
observed spectra.

\section{Experiment}

The material used in our experiments was purchased from Acros and
purified by gradient sublimation. This procedure yields small rubrene
crystallites which are relatively inert against oxidation in air
\cite{Kaefer_pccp05}.  To avoid well-known thermochromic effects --
rubrene changes its color towards the red at elevated temperatures
\cite{Schonberg_1954_JotACS_76_4134} -- we took all spectra at
$T=25\,^\circ$C.

The absorption spectra of rubrene in acetone solution were recorded
using a UV-Vis spectrophotometer (Varian Cary 50). With an overall
reproducibility of the optical constants being on a 1\% level we
verified that for the chosen concentrations ($1.46 \times 10^{-5} -
1.15 \times 10^{-4}$\,mol/l) the absorbance data follow the
Lambert-Beer law. 
The optical properties of rubrene thin films were measured in
ultra-high vacuum using a spectroscopic rotating compensator
ellipsometer (Wollam M-2000) with a broad band 75\,W Xe lamp
($250-1000$\,nm) and CCD based detection system with a spectral
resolution of about 1.6\,nm.  The organic material was out-gassed in
vacuum for several days at $\sim 190\,^\circ$C in order to remove
residual rubrene peroxide \cite{Wasserman_1971_JotAChS_94_4991}. Well
cleaned Si(100) substrates with a thermal oxide layer (thickness $\sim
20$\,nm) were transferred into the growth chamber and kept at
temperatures above $400\,^\circ$C for several hours. Before deposition
of the organic material we determined the thickness of the silicon
oxide and the precise angle of incidence by ellipsometry. The rubrene
films were deposited at room temperature via sublimation from a
Knudsen cell at $210\,^{\circ}$C. A typical deposition rate of
0.2\,nm/min and a pressure of less than $1 \times 10^{-9}$\,mbar
during growth ensured reproducible results.

\begin{figure}[htbp]
  \centering
  \includegraphics[width=0.8\columnwidth]{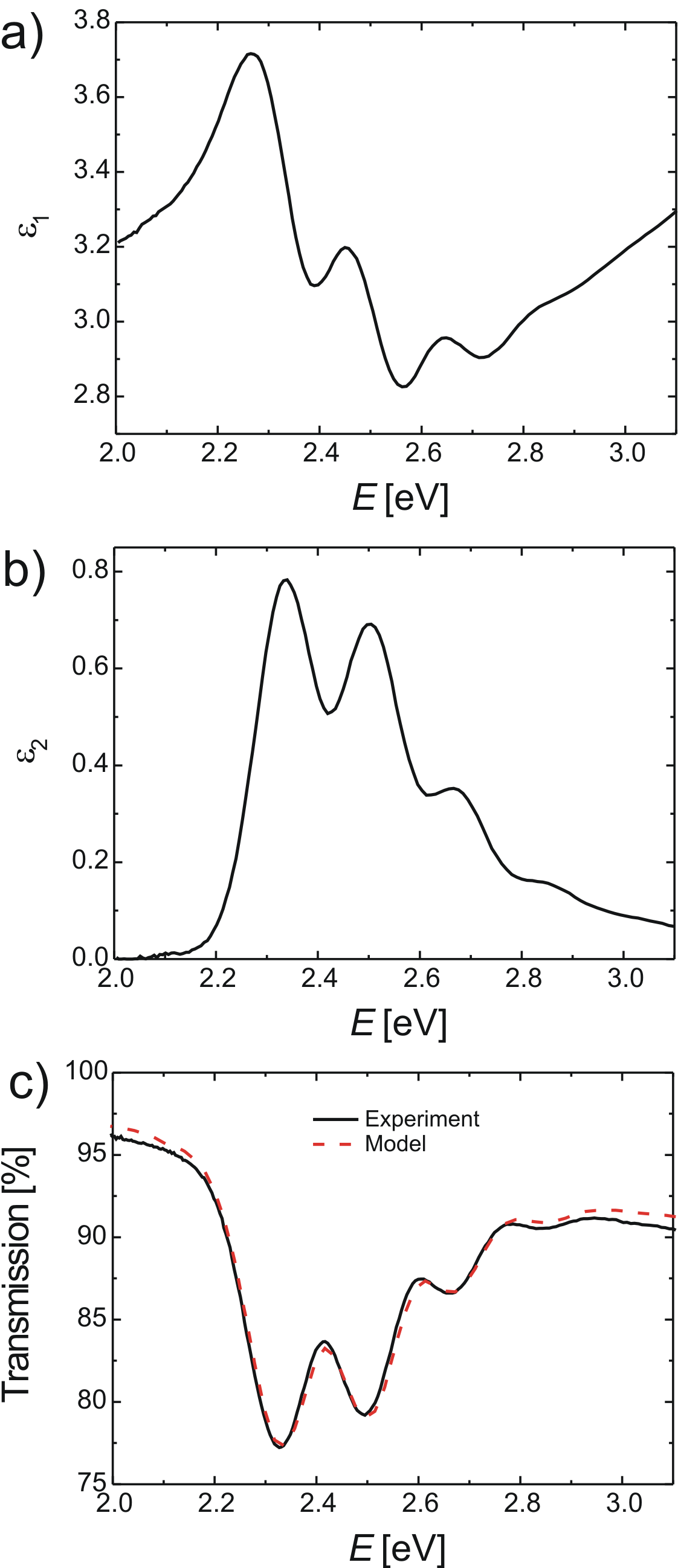}
  \caption{(Color online) Complex dielectric function of a 33\,nm
    thick rubrene film on SiO$_2$. (a) Real part $\varepsilon_1$ and
    (b) Imaginary part $\varepsilon_2$. The optical properties were
    obtained \emph{in situ} by spectroscopic ellipsometry and analyzed
    using a point-by-point fit. (c) Light transmission through a
    34\,nm thick rubrene film on glass measured in air (solid black).
    A model, which is based on the optical constants of rubrene shown
    above (a, b) and the optical constants of glass obtained
    separately by variable angle spectroscopic ellipsometry, can be
    used to describe the data (dashed red).}
  \label{fig:Film}
\end{figure}

The experimental data were analyzed with the commercial WVASE software
package \cite{Woollam_manual}. Since rubrene grows in amorphous films,
we used an isotropic model consisting of three layers, i.e.\ rubrene
-- silicon dioxide -- silicon, with the optical constants of silicon
and silicon dioxide taken from
Ref.~\cite{Herzinger_1998_JoAP_83_3323}.  First, we determined the
thickness of the rubrene film by employing a Cauchy model in the
spectral range below 2\,eV, where the absorption of the organic
material is negligible. Then, we performed a point-by-point fit which
yields the complex dielectric constant $\varepsilon_1+i \varepsilon_2$
for each energy, see Fig.~\ref{fig:Film}(a,b).  To check the
Kramers-Kronig consistency of the optical constants we also analyzed
the experimental data with a general oscillator model. In this
approach the imaginary part of the dielectric function $\varepsilon_2$
is described by a sum of Gaussian peaks and the real part
$\varepsilon_1$ is derived by the Kramers-Kronig relation.  Generally,
we found excellent agreement of the results obtained either by
point-by-point fits or the general oscillator model.  
%
%
To check the consistency of the results obtained by UV-Vis
spectroscopy and spectroscopic ellipsometry we also measured
transmission spectra of rubrene on glass, see Fig.~\ref{fig:Film}c.
We found that these data can be described very well using the optical
constants of rubrene on silicon obtained by ellipsometry (and those of
glass measured separately with variable angle ellipsometry).  The
minor differences between the data and the model might be due to
oxidation of some rubrene molecules in air
\cite{Mitrofanov_prl06,Kytka_apl07}.

\section{Results}

\subsection{Analysis of the optical spectra}

\begin{figure}[tbp]
  \centering
  \includegraphics[width=0.8\columnwidth]{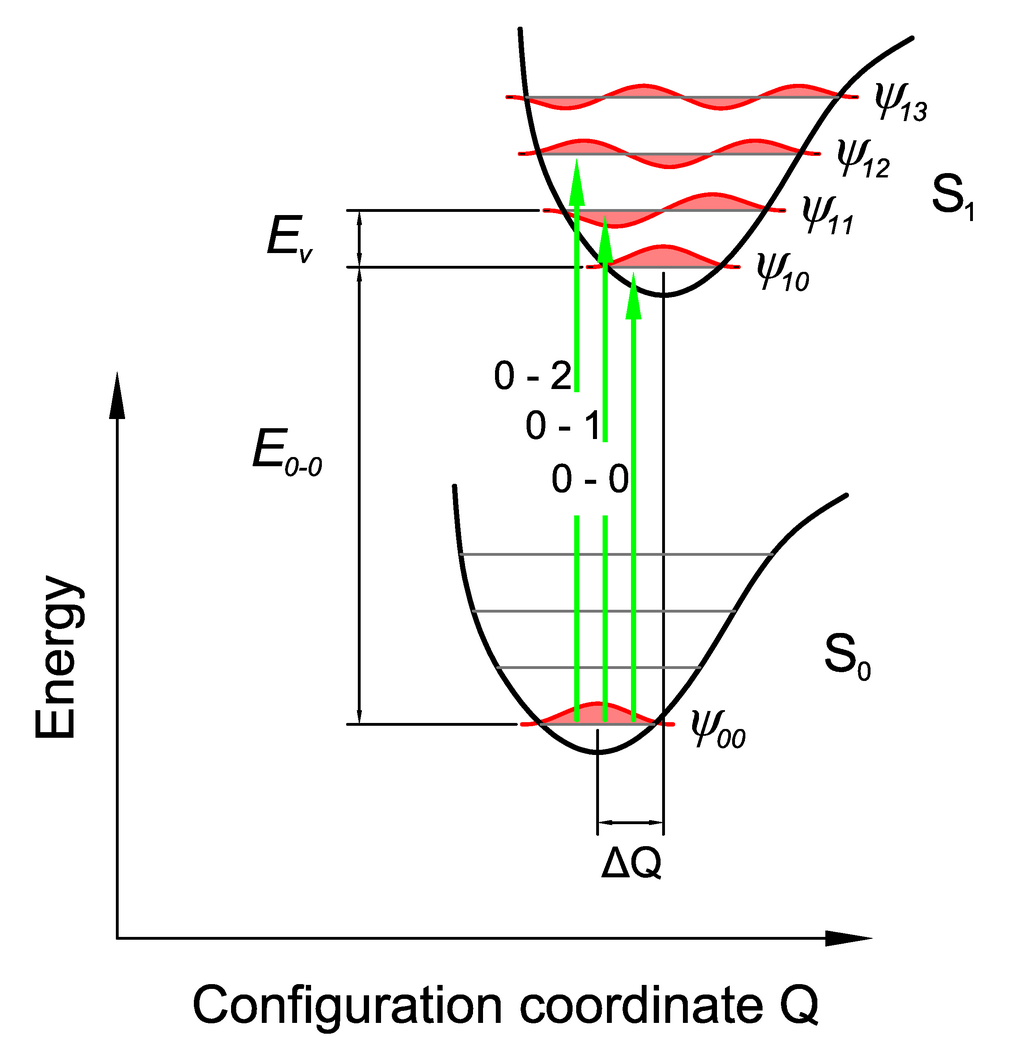}
  \caption{(Color online) Displaced harmonic oscillator model. 
    Because of different electron configurations in the $S_0$ (HOMO) and the $S_1$ 
    (LUMO) electronic excitations result in spatial deformations of 
    the molecule. The corresponding change of the 
    configurational coordinate $Q$ gives rise to the observed vibronic 
    excitations.%
  }
  \label{fig:DHO}
\end{figure}

The imaginary part of the dielectric function (Fig.~\ref{fig:Film}b)
shows a pronounced vibronic progression with five clearly
distinguishable sub-bands. The displaced harmonic oscillator model --
a simplified scheme describing the dipole-allowed transitions between
the electronic ground state ($S_0$) and the first excited state
($S_1$) including spatial deformations of the molecule
(Fig.~\ref{fig:DHO}) -- has been used to model the data.
In this approximation the excitation energies $E_n$ are given by 
\begin{equation}
  E_n= E_{0-0} + n E_v \ , 
  \label{eq:levels}
\end{equation}
where $E_{0-0}$ is the energy of the $S_{00} \rightarrow S_{10}$
transition, $E_v$ the vibrational energy and $n$ the vibrational
quantum number of the excited state.  The intensities $I_n$ of the
vibronic sub-bands contribute to the imaginary part of the dielectric
function according to a Poisson progression
\begin{equation}
  I_n = I \frac{S^n}{n!} e^{-S} \ ,
 \label{eq:HR_factor}
\end{equation}
where $S$ is the electron-phonon coupling constant, also known as the
Huang-Rhys factor. This coupling parameter essentially determines the
shape of the absorption spectrum, i.e.\ the relative intensities $I_n$
of the vibronic sub-bands, which are given by the area under each peak.

\begin{figure}[htbp]
  \centering
  \includegraphics[width=0.9\columnwidth]{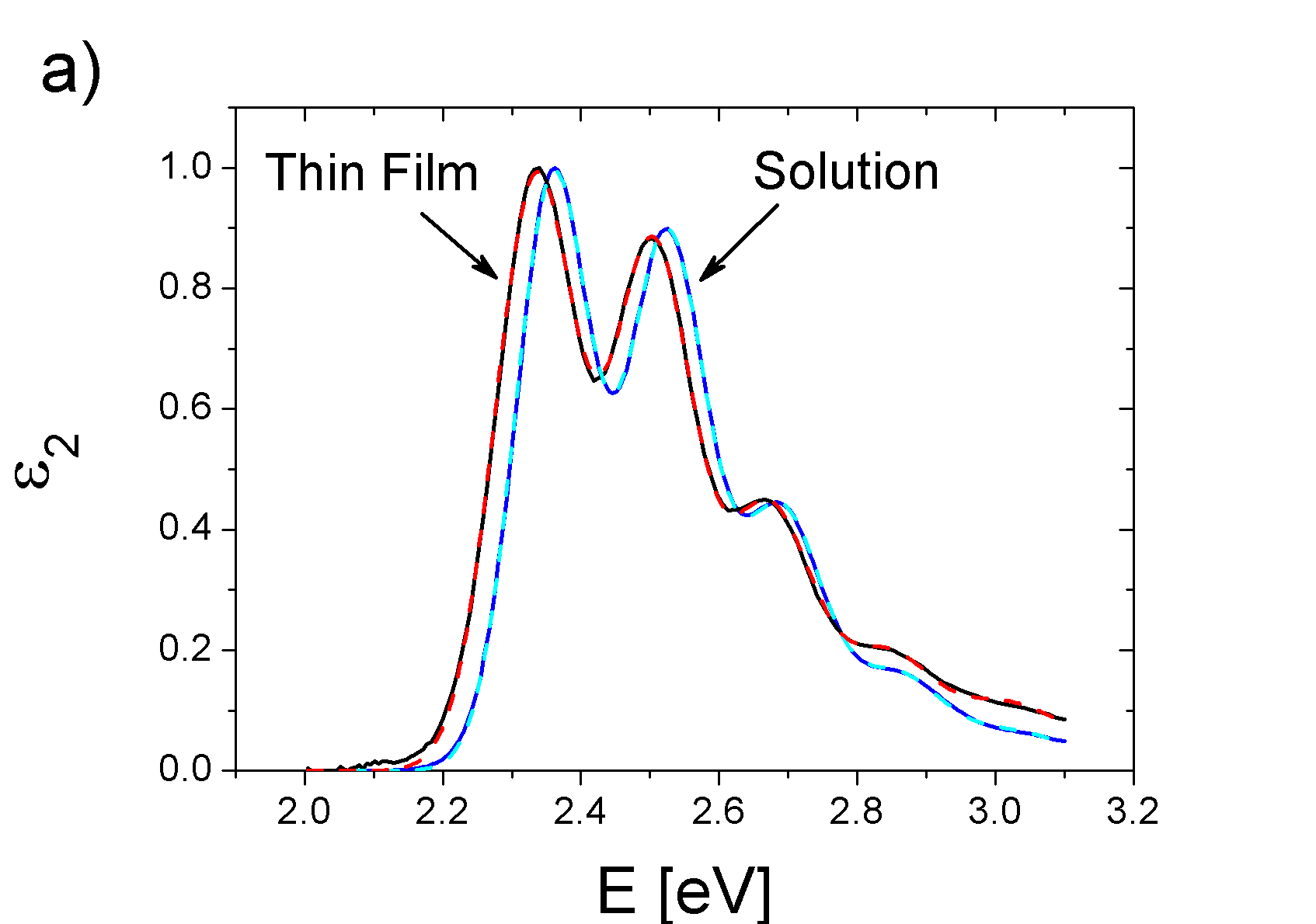}\\
  \includegraphics[width=0.9\columnwidth]{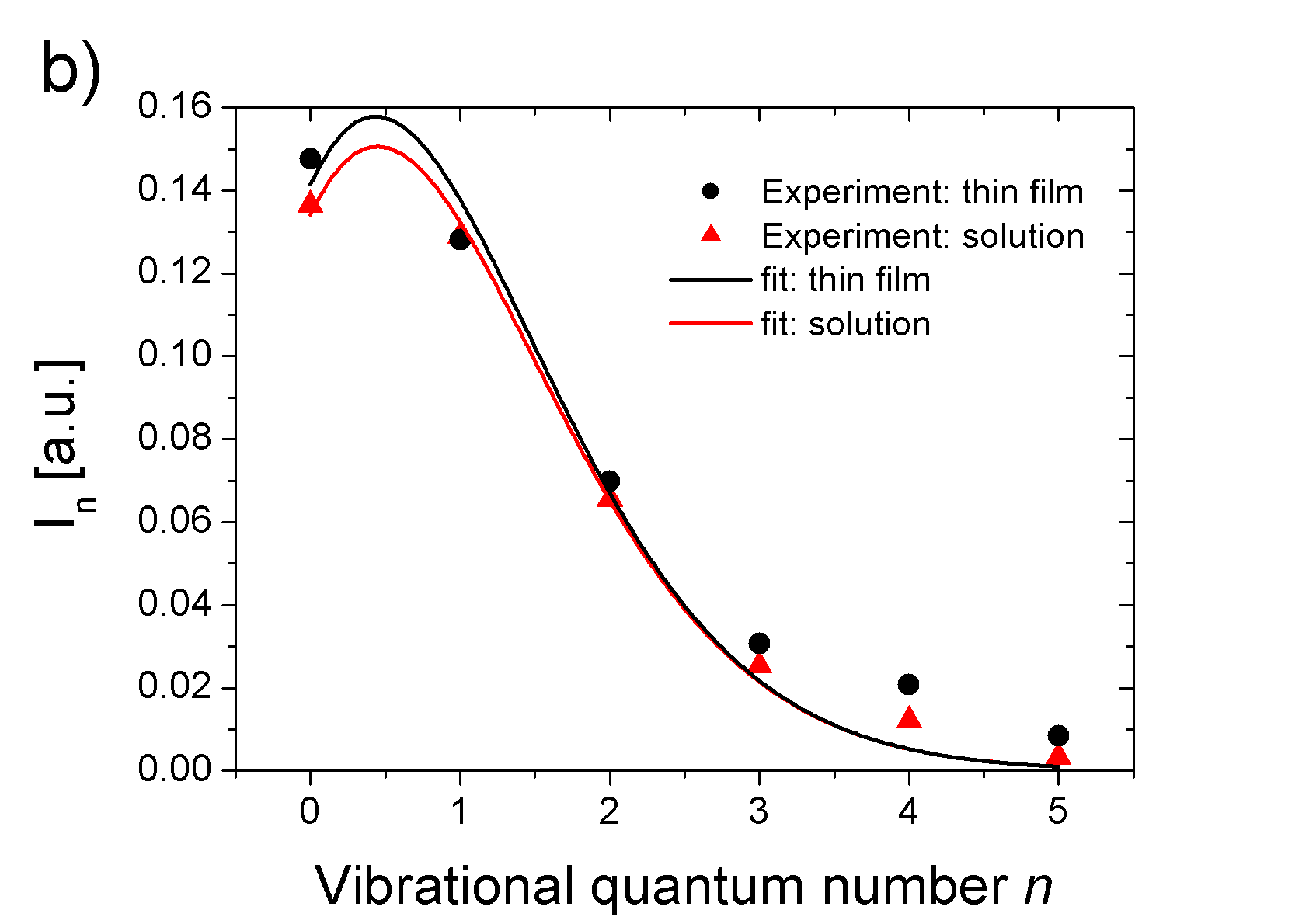}\\
  \includegraphics[width=0.9\columnwidth]{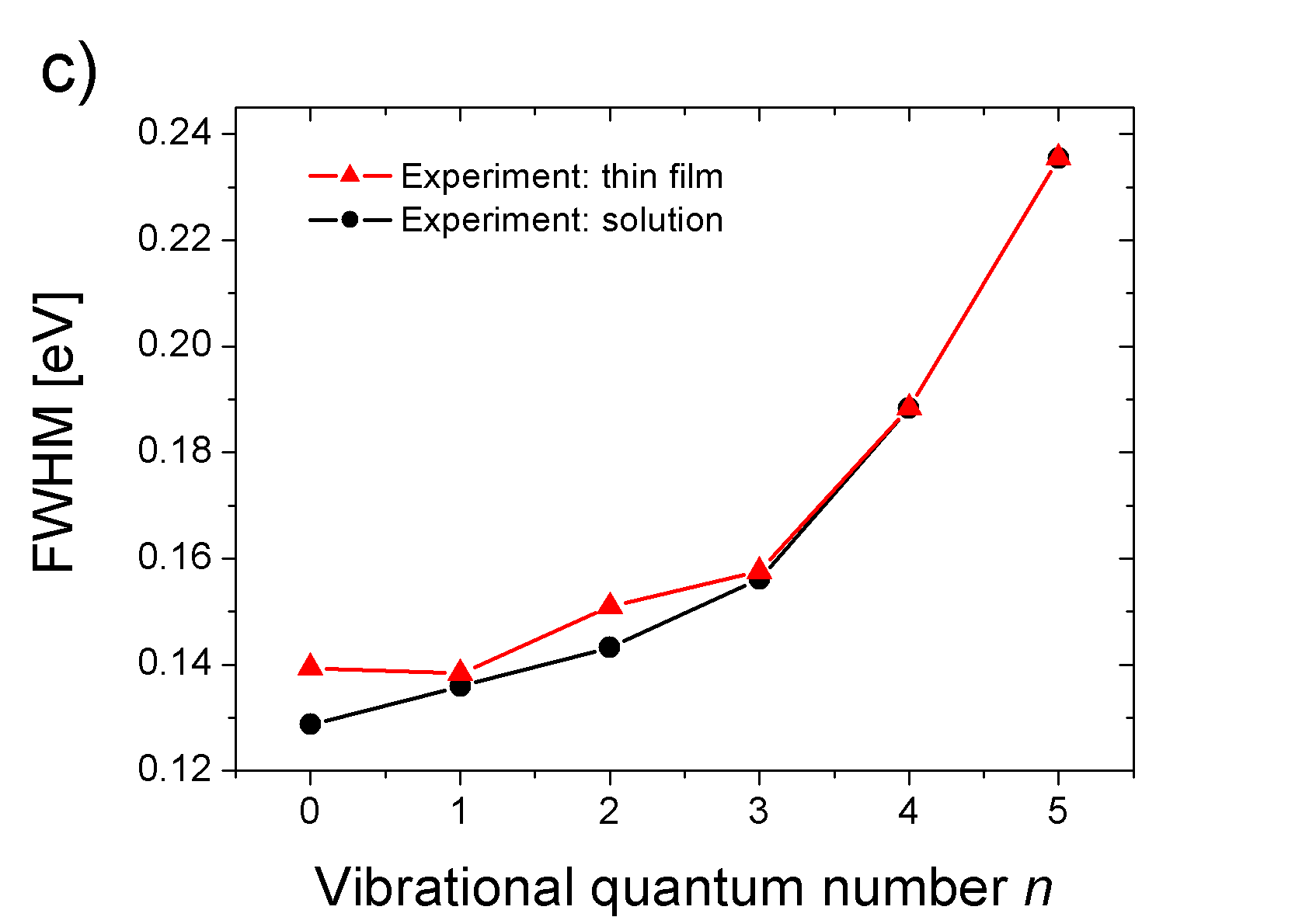}
  \caption{(Color online) (a) Normalized dielectric function
    $\varepsilon_2$ of a 33\,nm thick rubrene film (solid black) and
    of a dilute solution of rubrene in acetone (solid dark blue) with
    a fit to the data (dashed, red and light blue) based on
    Eq.~(\ref{eq:oscillator_model}).  The absorption spectrum shows an
    redshift which is caused primarily by the refractive index of the
    environment ('solvent shift'). (b) Least-square fit to the
    intensities $I_n$ of the vibronic progression using an
    interpolation of Eq.~(\ref{eq:HR_factor}) which yields the
    Huang-Rhys factors $S$ for the thin film and the monomer. (c) The
    width (FWHM) of the transitions peaks. In case of the solution it
    shows a monotonic increase, which indicates that several
    vibrational modes contribute to each sub-band.}
\label{fig:Solution_fit}
\end{figure}

In order to compare the data taken in solution with the thin film
spectra the absorption coefficient $\alpha$ of the solution has been
converted to the dielectric function using the common approximation
that the refractive index is virtually constant towards high energies
and $\varepsilon_2$ is then proportional to $\alpha / E$.
Fig.~\ref{fig:Solution_fit}a shows the normalized data for rubrene in
acetone solution and thin films together with the corresponding
least-square fits.

With the sub-band energies given by Eq.~(\ref{eq:levels}), their
relative intensities $I_n$ according to the Poisson progression in
Eq.~(\ref{eq:HR_factor}), and a Gaussian for each sub-band, the
imaginary part of the dielectric response can be expressed as
\begin{equation}
  \varepsilon_2 (E) = \sum_{n=0}^5  \frac{I_n}{\sigma_n \sqrt{2\pi}}  
  \exp \left[ - \frac{(E - E_n)^2}{2 \sigma_n^2} \right],
  \label{eq:oscillator_model}
\end{equation}
where the broadening parameters $\sigma_n$ are related to the
respective widths of each sub-band by $\mathrm{FWHM}_n=\sqrt{8 \ln
  2}\, \sigma_n$. Since several internal modes might contribute to
each sub-band, the linewidth are not necessarily constant.  In the
fitting algorithm all parameters except the width of the last two
peaks (fixed at $\sigma_4=0.08$\,eV and $\sigma_5=0.1$\,eV) have been
varied. The Huang-Rhys factor $S$ is then deduced from the intensities
of the first four transition peaks using Eq.~(\ref{eq:HR_factor}), see
Fig.~\ref{fig:Solution_fit}b \footnote{In principle, the Huang-Rhys
  parameter can be derived already from the intensities of only two
  adjacent peaks. Our approach, however, is more robust and therefore
  gives reliable results.}.
The results of the fitting procedure for the vibronic progression are
summarized in Tab.~\ref{tab:Experiment}. 
\begin{table}[h]
  \centering
  \begin{ruledtabular}
    \begin{tabular}{l|D{.}{.}{1.4}D{.}{.}{1.4}D{.}{.}{1.4}}
      & E_{0-0} & E_v & S \\
      \hline
      Solution &  2.359 & 0.165 & 0.986 \\
      Thin Film & 2.335 & 0.169 & 0.974 \\
    \end{tabular}
  \end{ruledtabular}
  \caption{Parameters derived from the experimental data using
    the model of the displaced harmonic oscillator for rubrene 
    in acetone solution and thin films: transition energy between the
    vibrational ground states $E_{0-0}$ (eV), vibrational energy
    $E_v$ (eV) and Huang-Rhys factor $S$.}
\label{tab:Experiment}
\end{table}
%



\subsection{DFT calculations}

In order to study the effect of possible conformational changes of the
molecule and to compare the experimental results with theory we have
computed the excitation energies and the deformation in the lowest
excited state of rubrene using the \textsc{turbomole 5.7} program
package \cite{Ahlrichs_cpl89}. We have chosen the B3LYP hybrid
functional since it gives excellent molecular geometries of organic
molecules together with rather precise transition energies resulting
from a compensation between the systematic deviations arising from the
underlying gradient-corrected density functional and the Hartree-Fock
contribution \cite{Bauschlicher_cpl95}. For the electronic orbitals,
we use a triple-$\zeta$ valence polarized (TZVP) basis
\cite{Schaefer_jcp94} throughout this article.

\begin{figure}[htbp]
  \centering
  \includegraphics[width=\columnwidth]{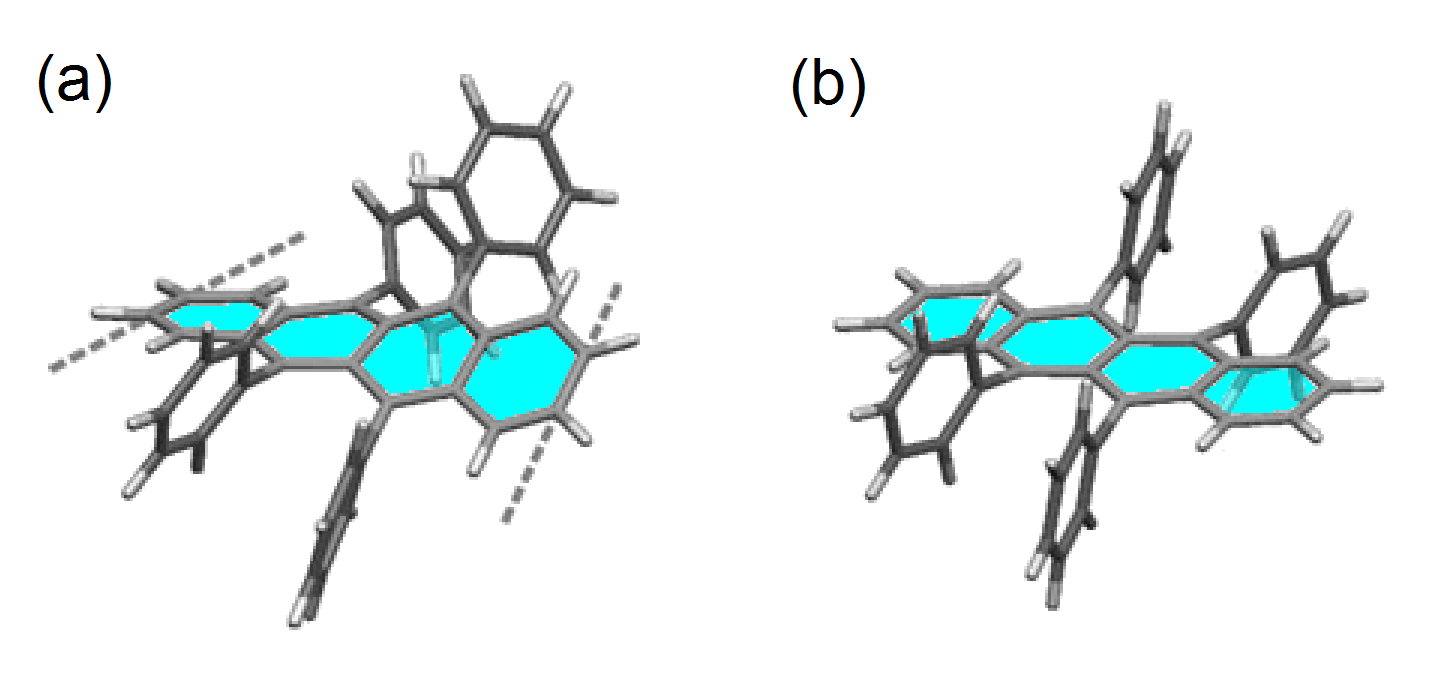}
  \caption{(Color online) (a) Geometry of the most stable rubrene
    isomer with a twisted tetracene backbone, and (b) planar rubrene
    isomer, which resembles the geometry in the crystalline phase.}
  \label{fig:isomers}
\end{figure}

In order to cover the twisted and the planar isomers of rubrene, we
have optimized both of them in their respective point groups, i.e.\
$D_2$ for the twisted isomer, and $C_{2h}$ for the planar one, see
Fig.~\ref{fig:isomers}.  Our simulation reproduces the known fact that
the twisted isomer is more stable \cite{Kaefer_prl05}, with an
energetic difference of $0.172$~eV at the B3LYP/TZVP level.

In solution the charge distribution of the solute molecule polarizes
the medium surrounding it, resulting in a further stabilization
energy.  This phenomenon can be handled with the conductor-like
screening model (COSMO), where the solute is treated as lying inside a
cavity surrounded by the polarizable medium \cite{cosmo}. In this
approach, the response of the dielectric medium is replaced by an
ideal conductor, generating screening charges on the surface of the
cavity. In a second step these screening charges are scaled by a
factor $f$ depending on the dielectric constant $\varepsilon$ of the
solvent, $f(\varepsilon)=(\varepsilon -1)/(\varepsilon +
\frac{1}{2})$. We have applied the respective routine as implemented
in the \textsc{turbomole 5.7} program package to a geometry
optimization of each isomer in a medium with a dielectric constant of
epsilon=20.7 corresponding to acetone. At the B3LYP/TZVP level we
found stabilization energies of $0.312$~eV for the planar isomer, and
$0.303$~eV for the twisted isomer. This is reducing the energetic
difference between the two isomers by a small amount, but the twisted
isomer still remains more stable by $0.163$~eV.

Even though the twisted isomer is more stable both as a free molecule
and in solution, in the crystalline phase the planar isomer can be
stabilized by a beneficial geometric arrangement between neighboring
molecules, resulting in a large cohesive energy so that the energetic
cost of the planarization is overcompensated by attractive
inter-molecular interactions. In the amorphous phase of solid rubrene
it is expected that the twisted geometry is conserved because the
random positions and orientations of the neighboring molecules cannot
stabilize an unfavorable isomer.

The different energies of both isomers raise the question if the
planar isomer is stable, or if it can spontaneously transform into the
more favorable twisted geometry. Using B3LYP and basis sets in the
size range between 6-31G(d) and TZVP we found that the lowest
symmetry-breaking $A_u$ mode in the $C_{2h}$ point group received a
small imaginary frequency. For the larger TZVP basis this apparent
instability was found to depend on the integration grid: A calculation
using the standard $m3$ grid resulted in an imaginary frequency of $i
13$~cm$^{-1}$, but based on a denser $m4$ grid, this value decreased
towards $i 5$~cm$^{-1}$. Using a tighter convergence of
$10^{-8}$~Hartree, the energetic minimum along this imaginary mode
occurs at 12~cm$^{-1} = 1.5$~meV below the $D_2$-symmetric
isomer. However, independently of basis size and integration grid, for
larger deformations along this vibrational eigenvector, the shape of
the potential returns to an essentially parabolic minimum, with a tiny
distortion for small deformations. Hence, this ultra-soft or even
imaginary mode does not constitute a viable barrierless pathway from
the planar isomer to the twisted configuration. For the twisted isomer
we found positive frequencies for all vibrational modes, the lowest
being a breathing mode at 26~cm$^{-1}$.

In the twisted isomer the repulsion between the phenyl side groups
achieves a rather larger distance between them as opposed to the
planar isomer, where a relatively small distance between the phenyl
groups is enforced by the rigidity of the bond connecting each side
group to the tetracene core and by the more restrictive point group
$C_{2h}$. Therefore, the twisted isomer gains a substantial amount of
energy through a reduction of these repulsive interactions, allowing
eventually to invest a part of this energy into the unfavorable twist
of the tetracene backbone. The resulting angle between the two central
rings of the tetracene core is $22.8^\circ$ in the electronic ground
state, and the angle between the two final rings $42.0^\circ$, compare
Fig.~\ref{fig:isomers}b.  In the relaxed excited state these angles
increase to $26.7^\circ$ and $43.6^\circ$, respectively.

For both isomers the relaxed excited geometries are computed with
time-dependent density functional theory (TD-DFT) at the B3LYP/TZVP
level, conserving in each case the point group of the electronic
ground state. The resulting deformation pattern is projected onto the
symmetry-conserving breathing modes of the respective isomer in its
ground state, defining in turn a Huang-Rhys $S_j$ factor for each
vibration $\hbar\omega_j$.
In contrast to the lowest symmetry-breaking vibration of the planar
isomer the frequencies of the symmetry-conserving modes depend only
weakly on variational basis set and integration grid, with the largest
influence for the lowest breathing mode found at 19~cm$^{-1}$
(B3LYP/TZVP, $m3$) or 22~cm$^{-1}$ (B3LYP/TZVP,
$m4$). Correspondingly, we have determined similar values for the
Huang-Rhys factors obtained at these levels. Applying the projection
scheme to the deformation pattern of a cationic molecule, we have
reproduced the key features of previous calculations addressing the
Huang-Rhys factors in the planar isomer \cite{dasilvafilho2005}. Of
course, when compared to the cationic geometry, a neutral excitation
of a planar rubrene molecule produces a somewhat different set of
Huang-Rhys factors.

By averaging over vibrations in the range between 900 and
1800\,cm$^{-1}$ we define an \textit{effective mode} with a Huang-Rhys
factor $S_\mathrm{eff}=\sum_j S_j$, a reorganization energy
$\lambda_\mathrm{eff}=\sum_j S_j \hbar \omega_j$, and a mode energy of
$\hbar \omega_\mathrm{eff}=\lambda_\mathrm{eff}/S_\mathrm{eff}$.  The
harmonic approximation to the potential energy surface results in
slightly too large vibrational frequencies. In order to compensate for
this deviation we apply a scaling factor of 0.973 adequate for the
B3LYP functional to the vibrational frequencies
\cite{Scott_jpc96,Halls2001}, conserving however the reorganization
energies $\lambda_j$ assigned to each mode, so that the Huang-Rhys
factors are increased by a factor of $1.028=1/0.973$. The results of
this computational procedure are reported in Tab.~\ref{tab:Simulations}. 
We use two kinds of relaxed excited geometries: First, the result of
the TD-DFT geometry optimization, and second, a modified excited
geometry where changes of the twisting angle of each phenyl group
around the bond connecting it to the tetracene core are
eliminated. From similar investigations on other molecules containing
phenyl groups we expect that this procedure minimizes artifacts
arising from finite angle effects related to the relatively large
changes of these angles in the relaxed excited geometry
\cite{scholz09}.

When basing our projection directly onto the TD-DFT geometry
for the excited state, the sum over the reorganization energies
assigned to each mode exceeds the available reorganization energies
$\lambda^{(g)}$ and $\lambda^{(e)}$ deduced from the potential energy
surfaces of ground and excited state, respectively. Using instead the
modified excited state geometry where the orientation of the phenyl
groups is adjusted to the ground state geometry, this artifact
disappears, indicating that the non-orthogonality problems arising
from the modified orientation of the phenyl rings are eliminated.
Moreover, in the second projection scheme the frequency assigned to
the effective mode is in better agreement with the observations.

\begin{table}[h!]
  \centering
  \begin{ruledtabular}
  \begin{tabular}{l|D{.}{.}{1.4}D{.}{.}{1.4}} 
    & \multicolumn{1}{c}{twisted}   & \multicolumn{1}{c}{planar} \\[1ex] \hline
    $E_\mathrm{abs}$                          & 2.173   & 2.279    \\ 
    $E_{0-0}$                             & 1.965   & 2.085    \\     
    $\lambda^{(g)}$                        & 0.204   & 0.195    \\ 
    $\lambda^{(e)}$                        & 0.208   & 0.194    \\ 
    $\hbar \omega_\mathrm{eff}$ (TD-DFT)      & 0.155   & 0.150    \\ 
    $S_\mathrm{eff}$ (TD-DFT)                & 1.216   & 1.358    \\ 
    $\hbar \omega_\mathrm{eff}$ (no twisting) & 0.162   & 0.158    \\ 
    $S_\mathrm{eff}$ (no twisting)            & 0.985   & 1.025    \\ 
  \end{tabular}
  \end{ruledtabular}
  \caption{Vertical transition energy $E_\mathrm{abs}$ (eV) in the ground state 
    geometry, transition energy between lowest vibrational levels $E_{0-0}$ (eV), 
    reorganization energies $\lambda^{(g)}$ (eV) and $\lambda^{(e)}$ (eV) on ground 
    and excited state potential energy surfaces, respectively, effective mode 
    $\hbar \omega_\mathrm{eff}$ (eV), and effective Huang-Rhys factor for the 
    twisted and planar isomers of rubrene, before (TD-DFT) and after (no twisting) 
    eliminating the influence of a modified orientation of the phenyl side groups 
    in the relaxed excited geometry.  All B3LYP vibrational frequencies are scaled 
    down by a factor of 0.973, and the Huang-Rhys factors $S$ are scaled up 
    correspondingly by a factor of $1/0.973=1.028$, so that the reorganization 
    energies are conserved.  All calculations have been performed at the B3LYP/TZVP level.}
  \label{tab:Simulations}
\end{table}

\section{Discussion}

The spectra of rubrene in solution and rubrene thin films show a
pronounced vibronic progression that we have analyzed using the
displaced harmonic oscillator model. For evaluation of the results the
parameters compiled in Tab.~\ref{tab:Experiment} should be compared
with each other and with previous investigations: We find that the
transition energy $E_{0-0} = 2.359$~eV of rubrene in acetone solution
deviates less than $2$~meV from a value reported earlier
\cite{Badger_1951_SA_4_280}.  Based on the systematic study of solvent
shifts presented in that work one can estimate that the HOMO-LUMO
transition of rubrene dissolved in acetone is red-shifted by about
$0.100$~eV with respect to a free molecule.  Consequently, the lowest
vibronic sub-band in the free molecule would be found at $2.459$~eV.
This illustrates that the transition energy $E_{0-0} = 2.335$~eV
measured for rubrene in thin films shows a significant redshift which
is caused by the environment.
Since the Huang-Rhys factors for rubrene monomers dissolved in acetone
and amorphous thin films show only minor differences, i.e.\
$S_\mathrm{sol} = 0.986$ and $S_\mathrm{film} = 0.974$, we conclude
that the exciton-phonon coupling is not markedly influenced by
aggregation of the molecules -- most likely because the investigated
thin films are amorphous and the phenyl side groups keep those
neighboring tetracene backbones apart that are mainly involved in the
$\pi \rightarrow \pi^*$ transition.

Furthermore, we observe a continuous increase of the peak width in the
vibronic progression of rubrene which indicates that more than one
vibrational mode is involved. For the solution spectrum we find a
simple monotonic dependency, whereas the thin film spectrum shows a
different behavior with the width of the first peak being larger than
the second one. One could speculate that this observation is related
to the reported existence of a conformational rubrene isomer with a
slightly different electronic transition energy \cite{Kaefer_prl05}.
In order to verify whether two different isomers contribute to the
optical spectra one might compare the experimental data with the
results of the DFT calculations for the twisted and planar molecule
(Tab.~\ref{tab:Simulations}): Despite the excellent agreement of the
calculated Huang-Rhys factors with our experimental
values~\footnote{The fact that the calculated Huang-Rhys factor for
  the twisted isomer nearly coincides with the observed value is
  accidental: For other applications of the same functional and basis
  set to the relaxed excited geometry of poly-aromatic molecules
  larger deviations are found.}  their small difference
($S_{\mathrm{twist}} = 0.985$, $S_{\mathrm{plan}} = 1.025$) cannot
provide the required sensitivity to to distinguish both isomers.  Our
calculations, however, show that the transition energies of the
twisted and planar monomer differ significantly with $\Delta E_{0-0} =
0.120$\,eV. Unlike spectra of the pure isomers a mixture of the two
molecules therefore would have to show some unusual spectral features.

\begin{figure}[htbp]
  \centering
  \includegraphics[width=0.8\columnwidth]{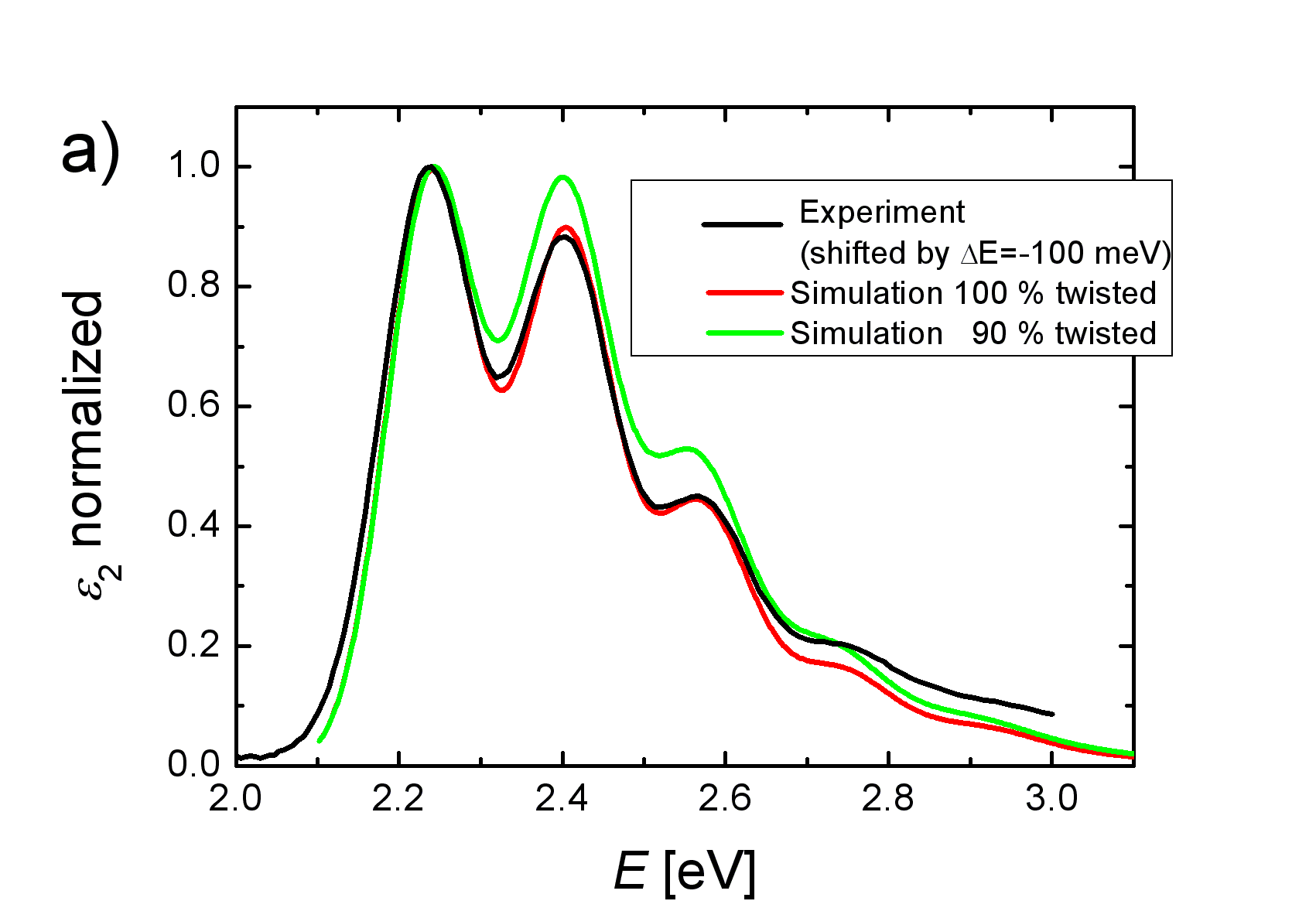}
  \includegraphics[width=0.8\columnwidth]{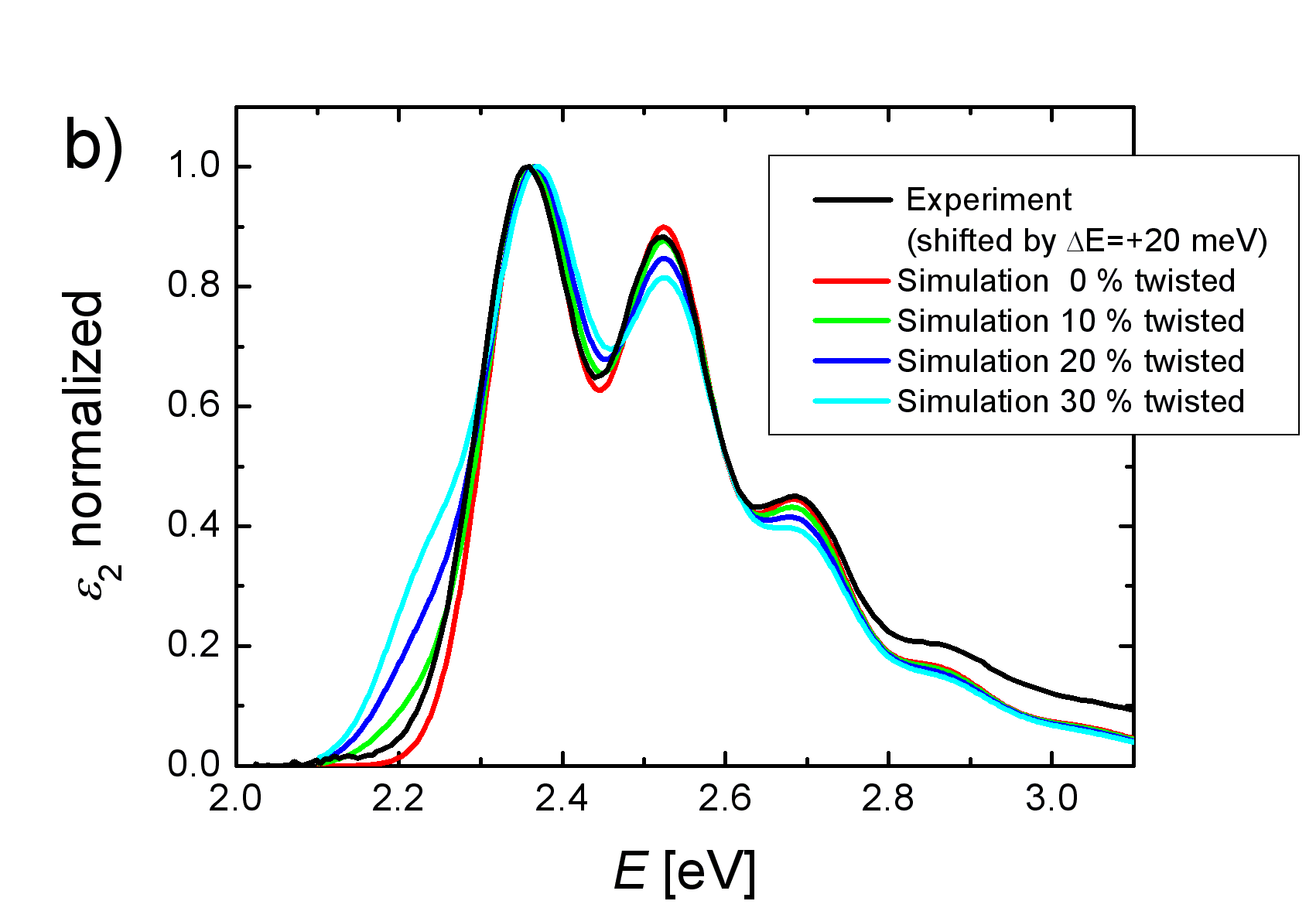}
  \caption{(Color online) Simulation of the spectra for different
    mixing ratios of planar and twisted rubrene isomers.  To
    compensate the solvent shift and to allow a comparison of the line
    shapes the thin film spectrum is shifted in energy such that the
    $E_{0-0}$ energy agrees with the simulated spectra. (a) mainly
    twisted isomers (b) mainly planar isomers.}
  \label{fig:Simulation_Mixing}
\end{figure}

Based on the solution spectrum of rubrene, which itself represents the
twisted isomer, we also model the spectrum of the planar isomer by
blue shifting the spectrum by $0.12$\,eV.  By adding both spectra with
different weights and normalizing them we simulate a corresponding
mixture of planar and twisted molecules, see
Fig.~\ref{fig:Simulation_Mixing}.
We find that the spectrum changes mainly in the low-energy range when
the number of twisted molecules is increased from e.g.\ 10 to 30\%
(Fig.~\ref{fig:Simulation_Mixing}b), whereas in the other case
increasing the number of planar molecules from 0 to 10\% affects
mainly the high-energy range (Fig.~\ref{fig:Simulation_Mixing}a).
Overall the simulations indicate that a contribution of the second
rubrene isomer is relative small, i.e.\ below 10\%.  Because the
twisted isomer is energetically favorable we conclude that the thin
film spectrum is dominated by the twisted molecule and a fraction of
no more than 10\% of planar molecules may contribute to the spectrum.
The differences between the solution and the thin film spectrum in the
high-energy range, however, cannot be explained by contributions of a
small fraction of planar molecules alone.  Additional coupling in the
thin film, like exciton transfer, might be the origin of this
deviation.

\section{Summary and Conclusions}

The optical spectra of rubrene thin films and monomers in solution
have been investigated experimentally by spectroscopic ellipsometry
and UV-Vis spectroscopy.  The analysis of the vibronic progression
using the displaced harmonic oscillator model has revealed the subtle
differences between the monomer and the thin film spectrum.
Additionally, time-dependent DFT calculations have been performed for
the planar and the twisted isomer in order to get more insight into
possible conformational changes of the molecule as they are known from
the literature.  The theoretical results, notably the excitation
energies and Huang-Rhys parameters for the planar and twisted
molecule, are in excellent agreement with the experimental data.
Since the two rubrene isomers show different transition energies, the
optical spectra may reveal their existence.  In fact a simulation,
which studies the effect of mixing the two species, indicates that one
isomer --~most likely the more stable twisted isomer~-- dominates the
thin film spectrum with a percentage of more than 90\%.

\section*{Acknowledgments}

The authors thank J.\ Pflaum for purifying rubrene, A.\ Mateasik for
making the UV-Vis spectrometer available and G.\ Witte for discussing
the results of our work.  We gratefully acknowledge financial support
by DFG, EPSRC, DAAD, Center of Excellence CENAMOST No.\ VVCE-0049-07
with support of project APVV-0290-06, VEGA 1/0689/09 and computational
resources at LRZ Garching.

\bibliographystyle{mystyle}


\end{document}